\documentclass[]{spie}  %

\usepackage{amsmath,amsfonts,amssymb}
\usepackage{graphicx}
\usepackage{booktabs}
\usepackage[colorlinks=true, allcolors=blue]{hyperref}

\title{Evaluating Cine and Tagged MRI-to-Speech Synthesis with Complementary
Metrics and MRI-Free Controls}

\author[a,b]{Karlo Vrancic}
\author[a]{Fangxu Xing}
\author[c]{Jiachen Zhuo}
\author[d]{Maureen Stone}
\author[e]{Jerry L. Prince}
\author[a]{Jonghye Woo}

\affil[a]{Dept. of Radiology, Mass General Brigham and Harvard Medical
School, Boston, MA, USA}
\affil[b]{Data Science, Harvard John A. Paulson School of Engineering and
Applied Sciences, Harvard University, Cambridge, MA, USA}
\affil[c]{Dept. of Radiology, University of Maryland School of Medicine,
Baltimore, MD, USA}
\affil[d]{Dept. of Neural and Pain Sciences, University of Maryland School of
Dentistry, Baltimore, MD, USA}
\affil[e]{Dept. of Electrical and Computer Engineering, Johns Hopkins
University, Baltimore, MD, USA}

\begin{document}
\maketitle

\begin{abstract}
Relating tongue motion measured with dynamic MRI to the speech it
produces could inform rehabilitation after glossectomy. Prior studies have translated cine or tagged
MRI into log-mel spectrograms and used the two-dimensional Pearson correlation
between predicted and recorded spectrograms (Corr2D) to quantify the acoustic
information captured by each imaging modality. However, Corr2D has not been
calibrated against MRI-free predictors that preserve basic acoustic structure.
Our contribution is an evaluation framework for speech synthesis from cine
MRI, tagged MRI and their combination; the synthesis network is
held fixed. On identical leave-one-subject-out folds, we
compare four trained input configurations with two MRI-free controls: a static
mean-spectrum baseline and a time-aligned cohort template. Across 44 subjects
(18 after glossectomy) producing two utterances, the cohort template, which never sees
an MRI, reached a Corr2D of 0.745 and 0.664. The static baseline, which
contains no timing, reached 87 to 89\% of that score. No trained configuration
outperformed the template on Corr2D, a cepstral distance or ESTOI; seven of
the eight fell significantly short on Corr2D. For cine
MRI this held
across three positional encodings, an encoder temporal-attention module and a
training objective matched to the metric. Cine and tagged MRI differed consistently within subjects, although
the direction depended on the metric. Absolute Corr2D alone
therefore does not show that a model recovers acoustics from the MRI. The two
controls are a reusable method: any study can compute them from its own
training recordings, so a model's margin over them measures what its images
contribute.
\end{abstract}


\section{INTRODUCTION}

Speech is produced by the coordinated deformation of the tongue and the
surrounding oropharyngeal muscles. Surgical treatment of oral cancer alters
that deformation directly. Patients who undergo glossectomy, the surgical
removal of part of the tongue, keep the intent to speak but lose part of the
articulator that realizes it. The resulting loss of intelligibility varies
widely between patients: consonant intelligibility correlates only moderately
with residual tongue motility, so how much speech a given resection will cost
cannot be predicted from the extent of the resection alone.\cite{bressmann2004}
Relating internal tongue motion, as measured by MRI, to the acoustic features
that determine intelligibility would give surgical planning and rehabilitation
the quantitative basis they currently lack.

Dynamic MRI is the measurement that makes this tractable \textit{in vivo}. Cine
MRI resolves tongue and oropharyngeal anatomy over time. Tagged MRI encodes
internal tissue deformation and supports dense motion
estimation.\cite{xing2017phase} Both have been used to build statistical
descriptions of speech motion in controls and
patients.\cite{woo2019speechmap,woo2021deep} To relate that motion to
acoustics, a line of work translates these sequences into two-dimensional (2D)
log-mel spectrograms as a proxy for the audio
waveform,\cite{liu2022cmri2spec,liu2022tagged,liu2023transformer,liu2023plt,li2025kecvae}
following the representation used for speech reconstruction from silent
video.\cite{akbari2018lip2audspec} The accuracy of the recovered spectrogram is
then read as a measure of how much of the acoustic signal internal motion
determines.

This reading has relied largely on the 2D Pearson correlation between predicted
and recorded log-mel spectrograms, referred to as Corr2D. Reported values range
from 0.665 to 0.820, across tagged-MRI translators, a 3D CNN with a temporal
module and its application to
patients.\cite{liu2022tagged,liu2023transformer,liu2024anomaly} These absolute
values have not been calibrated against predictions made without MRI. Without
such a reference, a score cannot be split into the part the imaging contributes
and the part any predictor would obtain. That split is what the clinical
question requires.

\begin{figure}[t]
\begin{center}
\includegraphics[width=1\linewidth]{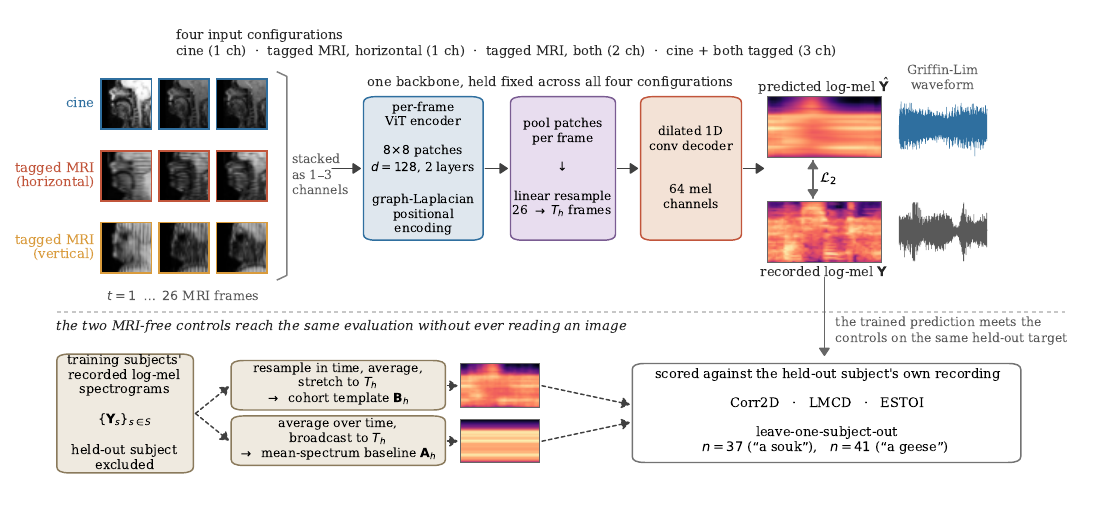}
\end{center}
\caption{The image-to-spectrogram network and the two MRI-free controls.
\emph{Top:} one backbone is held fixed across all four input configurations,
which differ only in how many image channels are stacked. Per-frame patch tokens
carry a graph-Laplacian positional encoding. They are pooled to one vector per
frame and resampled along time to the target length; a decoder then emits 64 mel
channels. \emph{Bottom:} the controls reach the same evaluation from the
training subjects' recorded spectrograms alone, without MRI. Both controls and
the trained models receive the held-out subject's duration. MRI frames,
spectrograms and waveforms are real, from the held-out subject of
Fig.~\ref{fig:qual}.}
\label{fig:arch}
\end{figure}

A random or chance-level predictor is not a sufficient reference here, for two
reasons. First, the cohorts in this line of work are built around one or two
short scripted utterances spoken to a metronome. An MRI-free acoustic template
can therefore score high by reproducing the shared structure of the utterance
alone. Second, Corr2D is a single Pearson correlation over a flattened log-mel
spectrogram. It ignores any positive scaling and offset of the prediction, while
log-mel speech carries a strong static spectral tilt common to every subject and
every frame. Part of a reported score is therefore a property of the
representation rather than of the model or the imaging. A useful reference has
to measure the utterance term and the representation term separately.

\begin{figure}[t]
\begin{center}
\includegraphics[width=1\linewidth]{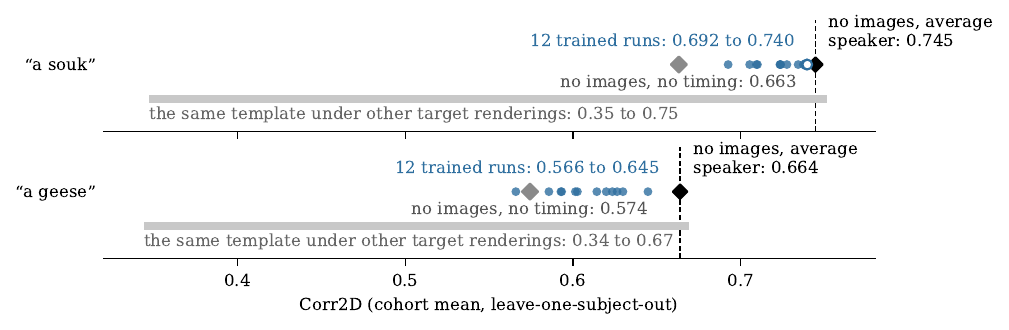}
\end{center}
\caption{Every primary-protocol Corr2D in this paper, one axis per utterance.
Grey diamond: the static mean-spectrum baseline (no images, no timing). Black
diamond and dashed line: the cohort template (no images; the average speaker).
Blue dots: the twelve trained runs of Sec.~\ref{sec:robust}, spanning
0.692--0.740 on ``a souk'' and 0.566--0.645 on ``a geese''. All lie below the
template in the mean; the two nearest (open circles), ``a souk'' cine under
either objective, are statistically level with it ($p=0.36$). Grey bar: the
template's Corr2D on the same recordings under other target renderings
(Sec.~\ref{sec:render}); these values are not comparable with the dots.}
\label{fig:ladder}
\end{figure}

We therefore construct two MRI-free predictors and evaluate them on the same
leave-one-subject-out folds as four trained input configurations: cine MRI,
single-orientation tagged MRI, two-orientation tagged MRI and the fusion of
cine with both tagged orientations. The first control, a static mean-spectrum
baseline, is the average spectrum of the training recordings, repeated over
time: it captures the overall spectral balance of the utterance but nothing
about the order or timing of its sounds. The second, a cohort template, is the
average of the training speakers' spectrograms of the same utterance, aligned
in time: it captures how a typical speaker produces the utterance. The only
thing it knows about the person being predicted is the length of the recording,
which the trained models also receive. Neither uses an image. Our central
finding is that the cohort template, which never sees an MRI, matches or
exceeds every model in this paper (Fig.~\ref{fig:ladder}). The contribution is
the evaluation, not the network: we hold one backbone fixed so that the inputs
and the controls are the only things that vary. Our contributions are: (i) the
two controls, with this finding on three metrics (Corr2D, a log-mel cepstral
distance and ESTOI) and the measurement that the mean-spectrum baseline, which
contains no timing, reaches 87--89\% of the template's Corr2D; (ii) evidence
that this result holds across three positional encodings, an encoder
temporal-attention module and a training objective matched to the metric's
invariance; (iii) evidence that the template's own score depends on how the
target is rendered, so it cannot be compared across studies; (iv) paired
within-subject modality comparisons and a patient-versus-healthy analysis, each
read in light of the controls.

\section{METHODS}

\subsection{Data}
\label{sec:data}

Paired cine MRI, tagged MRI in two orientations and audio were acquired on a
Siemens 3.0T TIM Trio or PRISMA with a segmented gradient echo
sequence.\cite{lee2013semi} Participants spoke ``a souk'' or ``a geese'' to a
metronome: 44 subjects (26 healthy speakers and 18 patients after glossectomy),
36 of whom performed both tasks. Because most subjects performed both tasks,
the two utterances are not independent replications, so we report them
separately rather than pooling them. Images are used at their native
$128\times128$ resolution, nominally with 26 frames per utterance. Audio is
converted to a 64-channel log-mel spectrogram at 22.05\,kHz (1024-point FFT,
256-sample hop), giving 55--123 frames (0.6--1.4\,s).

One recording per utterance is excluded from evaluation (Sub~1 for ``a souk''
and Sub~2 for ``a geese''), because the only available audio was an untrimmed
scan-session recording of 17.2\,s and 8.1\,s rather than the $\sim$1\,s
utterance used for every other subject. This leaves $n=37$ and $n=41$. Both
recordings remain in the training data of every other fold, for the controls
and the trained models alike. The exclusion changes no
template-versus-configuration conclusion: the template moves by about 0.003
Corr2D, while none of the 24 template-versus-configuration tests changes
significance. The one exploratory contrast it affects is reported in
Sec.~\ref{sec:patients}.

\subsection{Network and training protocol}

One image-to-spectrogram network is used in all conditions
(Fig.~\ref{fig:arch}). A per-frame vision transformer (two layers, four heads)
encodes $8\times8$ patches to 128-dimensional tokens with a positional encoding
built from the eight smallest non-trivial eigenvectors of the pixel-grid graph
Laplacian through a sign-invariant network. Patch tokens are pooled to one
vector per frame and the frame sequence is linearly resampled to the target mel
length. A dilated 1D convolutional decoder then emits 64 mel channels. Training
minimizes the mean-squared error on the log-mel spectrogram. The four
conditions differ only in the number of input channels: cine MRI (1),
single-orientation tagged MRI (1), two-orientation tagged MRI (2) and cine MRI
plus both tagged MRI orientations (3). Fusion is early: the cine and tagged MRI
frames of a subject are spatially and temporally aligned, so they are stacked as
channels and mixed by the first patch-embedding layer. A late-fusion variant,
with one encoder per input and their features averaged per frame, also falls
significantly short of the template on both utterances (Corr2D 0.720 and 0.604,
$p\leq0.002$). Every fold trains a fresh model with the same random seed for 40
epochs, so the input configuration is the intended variable.

The backbone has no temporal mixing in the encoder, so it is simpler than the
Longformer temporal module of prior work;\cite{liu2023transformer}
Sec.~\ref{sec:robust} adds encoder temporal attention at the same protocol. In
five subject-tasks fewer than 26 MRI frames are available in some
configurations, because loading multiple tagged orientations keeps only the
DICOM instance numbers they share and a few acquisitions are short. Restricting
the analysis to the subjects with 26 frames in every configuration leaves the
cine-over-two-orientation result unchanged and significant on both utterances
($p<0.001$).

\subsection{MRI-free controls}

Let $\{\mathbf{Y}_s\}_{s \in S}$ be the training spectrograms for a fold and
$T_h$ the held-out length. The cohort template resamples each training
spectrogram along time to a common length, averages them and resamples the
average to $T_h$. The static mean-spectrum baseline averages each training
spectrogram over time first, broadcasting one 64-dimensional spectrum across
all $T_h$ frames:
\begin{equation}
\mathbf{B}_h = \mathcal{R}_{T_h}\!\Big(\tfrac{1}{|S|}\textstyle\sum_{s}
\mathcal{R}_{\bar{T}}(\mathbf{Y}_s)\Big),
\qquad
\mathbf{A}_h = \Big(\tfrac{1}{|S|}\textstyle\sum_{s}
\tfrac{1}{T_s}\textstyle\sum_{t}\mathbf{Y}_s[:,t]\Big)\mathbf{1}_{T_h}^{\!\top},
\end{equation}
with $\mathcal{R}$ linear interpolation along time and
$\bar{T}=\operatorname{median}_s T_s$. Neither control uses MRI;
$\mathbf{A}_h$ also contains no timing. The controls and the models receive the
same target duration.

\subsection{Metrics}

We report Corr2D; a log-mel cepstral distance (LMCD), which applies the
mel-cepstral distortion formula\cite{kubichek1993mcd} to coefficients 1--13 of a
discrete cosine transform of our 64-channel log-mel spectrogram; and
ESTOI,\cite{jensen2016estoi} an intelligibility measure. Because LMCD is computed
from the log-mel spectrogram rather than from a full mel-cepstral analysis, its
values are an order of magnitude above those familiar from mel-cepstral
distortion. We use it only to rank models evaluated on identical data. ESTOI and
STOI\cite{taal2011stoi} differ on our waveforms by a per-subject offset that
cancels in a within-subject difference. Supporting analyses use sharpness
(variance of the Laplacian) and SSIM.\cite{wang2004ssim}

ESTOI is computed from Griffin--Lim\cite{griffin1984signal} reconstructions of
both the predicted and the ground-truth mel spectrograms, which holds the
vocoder constant. Griffin--Lim applied to the ground-truth mel spectrogram
scores STOI above 0.87 against the original recording, so the vocoder is
unlikely to limit the comparison. Unseeded phase initialization moves a single
ESTOI value by about 0.02, so cohort means are precise to about 0.003 and the
paired tests are valid, while individual subject values are not. Corr2D and LMCD
are computed directly in mel space and are unaffected by phase initialization.

\subsection{Statistics}

Within-subject comparisons use the Wilcoxon signed-rank test; patient-versus-healthy
comparisons use the Mann--Whitney $U$ test. The eight
template-versus-configuration Corr2D tests form the confirmatory family. Every
significant result in it survives Holm correction; the single non-significant
one is ``a souk'' cine MRI. Tests outside this family are exploratory. Exact
$p$-values against the template are in Table~\ref{tab:main}.

\begin{figure}[t]
\begin{center}
\includegraphics[width=1\linewidth]{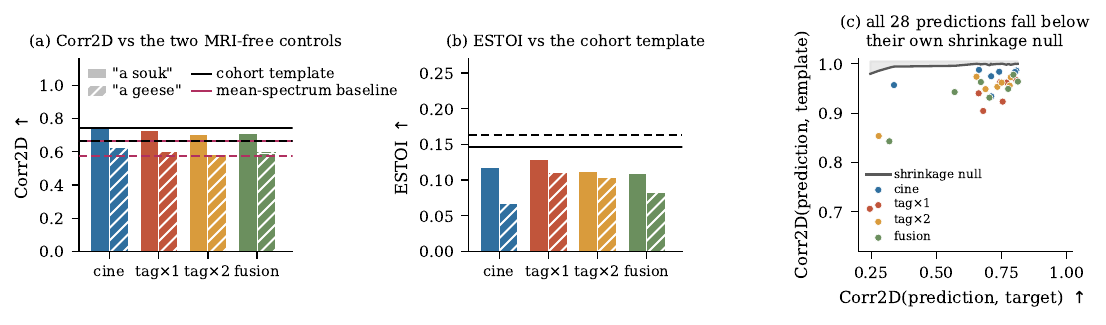}
\end{center}
\caption{(a,b) The four trained configurations against the two MRI-free
controls, solid fill for ``a souk'' and hatched for ``a geese''. No
configuration beats the template on either metric. On ``a geese'' only cine MRI
is significantly better than the static mean-spectrum baseline. (c) Each
prediction is compared with a matched reference $\alpha\mathbf{Y}+
(1-\alpha)\mathbf{B}$ whose fidelity to the target equals the prediction's own
(grey curve; shaded region is above it). All 28 predictions fall below their
matched reference. Points are 7 subject-tasks from a separate inference run
(Sec.~\ref{sec:collapse}).}
\label{fig:control}
\end{figure}

\section{RESULTS}

\subsection{The MRI-free controls}

Most of the template's score is available before any timing is modeled. The
static mean-spectrum baseline, one average spectrum repeated over time, reaches
87--89\% of the template's Corr2D on both utterances (Table~\ref{tab:main}).
Adding the time structure of the utterance supplies the rest, 0.081 on ``a
souk'' and 0.089 on ``a geese''. Under our rendering, a Corr2D of 0.57--0.66 is
therefore available before any modeling (Fig.~\ref{fig:ladder}). We do not
compare this value with published scores, because the controls themselves
depend on the rendering (Sec.~\ref{sec:render}) and must be measured within each
study.

\begin{table}[t]
\centering
\caption{Full leave-one-subject-out results with identical folds and training
protocol, against the two MRI-free controls. Parenthesized values are paired
Wilcoxon $p$ against the template. Best per column in \textbf{bold}. LMCD is
comparable across these rows only; LMCD and ESTOI were not computed for the
static baseline. The cohort template is best in every column; no configuration
exceeds it.}
\vspace{+5pt}
\label{tab:main}
\resizebox{1\linewidth}{!}{
\begin{tabular}{lccc c ccc}
\toprule
& \multicolumn{3}{c}{``a souk'' ($n=37$)} & & \multicolumn{3}{c}{``a geese'' ($n=41$)} \\
\cmidrule{2-4}\cmidrule{6-8}
Input & Corr2D $\uparrow$ & LMCD $\downarrow$ & ESTOI $\uparrow$ & &
        Corr2D $\uparrow$ & LMCD $\downarrow$ & ESTOI $\uparrow$ \\
\midrule
cine (1 ch)                  & 0.7393 \tiny{(0.36)} & 79.96 \tiny{(0.50)} & 0.1172 \tiny{(0.008)} & & 0.6262 \tiny{($<$.001)} & 88.19 \tiny{($<$.001)} & 0.0678 \tiny{($<$.001)} \\
tagged, 1 orientation (1 ch) & 0.7275 \tiny{(0.011)} & 83.45 \tiny{(0.005)} & 0.1289 \tiny{(0.21)} & & 0.6014 \tiny{($<$.001)} & 91.64 \tiny{($<$.001)} & 0.1102 \tiny{(0.002)} \\
tagged, 2 orientations (2 ch)& 0.7053 \tiny{($<$.001)} & 86.30 \tiny{($<$.001)} & 0.1117 \tiny{(0.017)} & & 0.5856 \tiny{($<$.001)} & 92.51 \tiny{($<$.001)} & 0.1037 \tiny{($<$.001)} \\
cine + tagged fusion (3 ch)  & 0.7099 \tiny{($<$.001)} & 85.69 \tiny{($<$.001)} & 0.1093 \tiny{(0.068)} & & 0.5927 \tiny{($<$.001)} & 91.91 \tiny{($<$.001)} & 0.0822 \tiny{($<$.001)} \\
\midrule
\emph{cohort template} & \textbf{0.7445} & \textbf{79.38} & \textbf{0.1466} & & \textbf{0.6638} & \textbf{84.12} & \textbf{0.1632} \\
\emph{static mean-spectrum baseline} & 0.6631 & -- & -- & & 0.5743 & -- & -- \\
\bottomrule
\end{tabular}}
\end{table}

\subsection{Trained configurations against the controls}

No trained configuration outperforms the cohort template (Table~\ref{tab:main},
Fig.~\ref{fig:control}a,b). On ``a geese'' every configuration falls
significantly short of it on all three metrics. On ``a souk'' three of the four
fall short on Corr2D. The fourth, cine MRI, ties the template on Corr2D and LMCD
but is worse on ESTOI. The tie does not hold up: the same configuration
retrained on different hardware falls significantly short of the template
($p=0.021$), while its ``a geese'' comparison replicates. We therefore read
cine MRI on ``a souk'' as at best level with the template.

Against the static baseline, which contains no timing at all, the
configurations separate by utterance. On ``a souk'' all four outperform it
significantly ($p\leq0.004$). On ``a geese'' all four exceed it in the mean, but
only cine MRI does so significantly ($p=0.0001$). The two tagged MRI configurations and fusion do not differ
significantly from a predictor with no temporal information.

\subsection{Robustness to the backbone}
\label{sec:robust}

Adding temporal attention to the encoder helps on the harder utterance but does
not close the gap to the template. One explanation for the template's advantage
is that the backbone, which has no temporal mixing in its encoder, is
underpowered in time. We therefore added two transformer layers attending over
frames in the encoder and retrained the cine MRI configuration with the full
leave-one-subject-out protocol. The module improves the model significantly on
``a geese'' ($p=0.033$) but not on ``a souk''. On both utterances it still falls
significantly short of the template ($p\leq0.021$).

The choice of positional encoding does not change the result either. Because the
graph-Laplacian encoding is an uncommon choice, we retrained the cine MRI
configuration with a learned per-patch embedding and with no spatial positional
information. All three encodings fall significantly short of the template on
both utterances ($p\leq0.021$). The learned embedding is the weakest, significantly
worse than no encoding at all ($p\leq0.014$), as expected from its 32{,}768 free parameters estimated from
about 40 subjects; on ``a geese'' it even drops slightly below the mean-spectrum
baseline. The graph-Laplacian encoding improves significantly on no encoding for
``a geese'' and matches it for ``a souk''. Both the temporal module and the
positional encoding therefore help on the harder utterance, while neither clears
the template.

Across all twelve primary-protocol runs, Corr2D separates the configurations
from one another but places none of them above a predictor that reads no images.
The twelve runs are the four input configurations under the mean-squared error
and under the $z$-scored objective of Sec.~\ref{sec:objective}, the two
alternative positional encodings, the graph-Laplacian rerun and encoder temporal
attention. Every one scores below the template in the mean on both utterances
(Fig.~\ref{fig:ladder}). Only ``a souk'' cine MRI, under either objective, is
level with it; every other run falls significantly short. On ``a souk'' the
whole spread of the trained runs is narrower than the 0.081 that separates the
template from the mean-spectrum baseline.

\subsection{Dependence on target parameterization}
\label{sec:render}

The template's own score depends more on how the audio is rendered than
published scores differ from one another. We rendered the same recordings to a
fixed $64\times64$ log-mel spectrogram in two ways: by time-warping the full
spectrogram to 64 frames, or by taking a fixed 21,000-sample window, the length
used for sliding-window augmentation in prior work on this
cohort.\cite{liu2023transformer} We also restricted the spectrogram to the
40--1000\,Hz band used in the same work. Warping raises the template slightly,
the fixed window lowers it and the restricted band lowers it most. An MRI-free
predictor therefore spans 0.34--0.75 on identical recordings depending only on
the rendering, a range more than 2.5 times wider than the 0.665--0.820 spread of
all published MRI-to-speech scores. Because some of our clips are shorter than
the window and are zero-padded, whereas the recordings in that work were
reported as longer than the window,\cite{liu2023transformer} our cropped
condition is not an emulation of any published pipeline. What it shows is
general: Corr2D values from different studies cannot be ranked against each
other, or read against a fixed threshold, without a control computed under each
study's own rendering.

\begin{figure}[t]
\begin{center}
\includegraphics[width=1\linewidth]{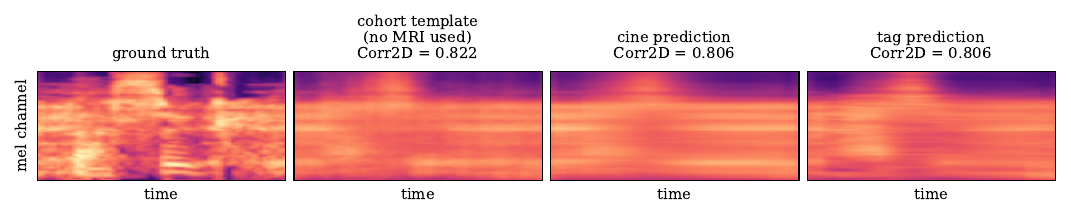}
\end{center}
\caption{One held-out healthy subject, ``a souk''. The recorded spectrogram
carries temporal events that none of the predictions recover. The cohort
template, the cine MRI prediction and the tagged MRI prediction are visually the
same object; the template scores the highest Corr2D of the three. This subject,
among the better-scoring folds, is shown for legibility. It comes from the same
separate inference run as Fig.~\ref{fig:control}c.}
\label{fig:qual}
\end{figure}

\subsection{Relation of the predictions to the template}
\label{sec:collapse}

The predictions stay close to the template and add noise rather than
subject-specific detail. For seven subject-tasks whose predictions were kept
from a separate inference run, a prediction correlates more with the cohort
template than with its own target (Corr2D 0.94 against 0.67 on average;
Fig.~\ref{fig:control}c). That contrast alone is weak evidence, because any
estimator shrunk toward the cohort mean satisfies
$\mathrm{Corr2D}(\hat{\mathbf{Y}},\mathbf{B})>\mathrm{Corr2D}(\hat{\mathbf{Y}},
\mathbf{Y})$, even one that carries subject-specific information. We therefore
built, for each prediction, a matched reference
$\alpha\mathbf{Y}+(1-\alpha)\mathbf{B}$ whose fidelity to the target equals the
prediction's. For 21 of the 28 predictions the matched $\alpha$ is negative: they are less
faithful to their target than the template itself. For the other seven it is at
most 0.04, a mixture of 4\% target and 96\% template. All 28 predictions
also depart from the template more than their matched references do
(Fig.~\ref{fig:control}c). The predictions were thus not simply shifted toward
the cohort mean: they stay close to the template and depart from it in
directions that do not improve fidelity. The seven subject-tasks were chosen for
listening rather than at random. Their values differ from the
Table~\ref{tab:main} run by up to 0.15 Corr2D, so we treat this analysis as
illustrative.

\subsection{Paired modality comparisons}

Within subjects, cine MRI outperforms two-orientation tagged MRI on Corr2D for
both utterances, but the ranking reverses on ESTOI for ``a geese''. Paired comparisons rank
configurations on identical targets. Cine MRI beats two-orientation tagged MRI on
Corr2D in 26 of 37 and 31 of 41 subjects ($p\leq0.0007$); both contrasts survive
Holm correction together with the confirmatory family. Cine MRI does not differ
significantly from single-orientation tagged MRI, while on ``a souk'' one tagged
orientation significantly outperforms two ($p=0.021$). Fusion does not outperform cine MRI:
it lands next to two-orientation tagged MRI, the input it mostly consists of.
With about 40 training subjects, each added channel gives the network more input
to fit without more subjects to fit it from. Because every configuration stays
close to the template, the extra input mostly adds variation around it rather
than information. On ESTOI the ordering reverses on ``a geese'': both tagged MRI
configurations significantly outperform cine MRI, the same pair that cine leads
on Corr2D.

These contrasts do not isolate MRI-derived information, because the
configurations differ in how closely they follow the template. In the seven-subject-task run of Sec.~\ref{sec:collapse}, cine MRI predictions correlate with the template at an average of 0.97, while single-orientation tagged MRI predictions correlate at 0.91.~A Corr2D advantage can therefore reflect stronger shrinkage
toward the template; the reversed ESTOI ordering fits this reading. The
controls change how the modality comparison reads: the configurations differ in
how they depart from an MRI-free predictor, with a direction that depends on the
metric.

\subsection{A training objective matched to the metric}
\label{sec:objective}

Corr2D misses a large change in what the predictions look like. It
ignores positive scaling and offset, so we also trained the four configurations
on the per-utterance $z$-scored log-mel spectrogram, with level supervised
separately. For cine MRI this makes predictions 1.6 to 1.9 times sharper in every
subject and improves SSIM significantly, while Corr2D changes by at most 0.004.
The sharpness has a cost: LMCD, PSNR and log-mel L2 all degrade significantly
(all $p<2\times10^{-5}$). Corr2D is thus
nearly blind both to a large sharpness gain and to a simultaneous loss of level
fidelity. This is the perception-distortion tradeoff\cite{blau2018perception} in
a setting where the distortion measure is itself insensitive to the shared
prior.

\subsection{Patients and healthy speakers}
\label{sec:patients}

Patients and healthy speakers were similarly predictable, contrary to the
expected clinical pattern. Because glossectomy alters both tongue motion and
speech acoustics, a model that had learned the articulatory-acoustic
relationship should find patients harder to predict. Performance was instead
similar for every configuration on both tasks: none of the eight Corr2D
contrasts reaches significance ($p\geq0.25$), with 23 healthy speakers against 14 and 18
patients. This is not model robustness. The MRI-free template itself scores
patients and healthy speakers equally ($p\geq0.52$), so under Corr2D patients are not harder
targets. A null contrast for the trained models therefore carries no
information about clinical generalizability. ESTOI showed a less consistent
pattern: on ``a souk'' single-orientation tagged MRI scored lower in patients
than in healthy speakers, short of significance ($p=0.058$, exploratory). This is the only contrast whose significance depends on the exclusion described in Sec.~\ref{sec:data}: if the excluded patient recording is retained, the result reaches significance ($p = 0.039$).

\section{DISCUSSION AND CONCLUSION}

The two controls split a Corr2D score into its parts. The mean spectrum
measures what the log-mel representation supplies. The template adds what the
shared utterance supplies. Only a model's margin over the template can come from
the individual speaker. The first part alone is 87--89\% of the template's
score, so the largest share is a property of log-mel speech rather than of the
corpus. The shrinkage analysis of Sec.~\ref{sec:collapse} finds at most a
marginal individual part: none of the 28 listening-subset predictions is more
faithful to its target than a mixture of 4\% target and 96\% template.

For the clinical question, this changes what existing scores support. A high
Corr2D invites the reading that internal tongue motion largely determines the
acoustics it produces. On our cohort and rendering, a predictor with no access
to motion reaches 0.745. An absolute Corr2D is therefore evidence about the
imaging only through its margin over the strongest MRI-free predictor evaluated
on the same data; in our data, no trained model has a positive margin. The
missing piece is evidence of subject-specific acoustic recovery: no current
score demonstrates it with the sensitivity a surgical-planning or rehabilitation
tool would need. The patient-versus-healthy result points the same way: neither
the models nor the metric and target, as currently defined, separate an impaired
speaker from an unimpaired one.

The claim is about evidence, not about any published model. This work uses the same cohort and scanner protocol as prior MRI-to-speech studies from our group. However, we have not run the control experiments under their target representation and training setup. Our backbone is also simpler. A stronger model may well clear
the template; only a measured margin over it can show that it does. The margin
is narrow at one point: ``a souk'' cine MRI ties the template under either objective
and becomes a significant shortfall on retraining. Every other run at the
primary protocol falls significantly short of the template, including runs with
a temporal module and a positional encoding that each improve significantly on
the backbone without them on ``a geese''.

These results suggest a short reporting checklist for studies that synthesize
speech, or another time-varying signal, from imaging in a cohort performing a
shared scripted task:

\begin{center}
\fbox{\parbox{0.94\linewidth}{\small
\textbf{Reporting checklist for image-to-signal synthesis}
\begin{enumerate}\itemsep 1pt
\item Report a static baseline: the training targets averaged over time (for
spectrograms, the mean spectrum). It requires no training and shows how much of
a score the representation alone supplies.
\item Report a cohort template on the same folds, durations and rendering. Quote
each model's margin over it with a paired test.
\item State the target rendering (for spectrograms: sampling rate, window,
frequency band, length normalization): an MRI-free predictor spans 0.34 to 0.75
on our recordings depending on it.
\item Compare input modalities within subject and on at least two metrics of
different kinds (for speech, one spectral and one intelligibility measure).
Report how closely each model follows the template, since a paired difference
can reflect shrinkage as well as information.
\end{enumerate}}}
\end{center}

This study has limitations. Two scripted utterances from one site form the
setting in which a template is strongest; the mean-spectrum result suggests that
a corpus with varied content\cite{lim2021multispeaker} would remove the smaller,
utterance-dependent part of the score rather than the larger one. Each input is
a single 2D slice per frame, so lateral deformation and rotation of the tongue
are not observed; the controls use no images and apply unchanged to 3D or
motion-field inputs, which must clear the same template. The two tasks share 36
of 44 subjects and are not independent. The audio clips were cut by hand from
the in-scanner recording and are not synchronized to the scanner trigger.
Against expert articulatory landmarks, the proportional frame-to-spectrogram
mapping used by the models is off by a median of 83\,ms on ``a souk'' and
120\,ms on ``a geese'', two to three MRI frames. Because each clip begins and
ends at the utterance, the template inherits the utterance boundaries, whereas
the models must map misaligned frames, so the template comparison favors the
control; the static-baseline result, which uses no timing, is unaffected. We use
Griffin--Lim rather than a neural vocoder.\cite{kong2020hifigan} Our ESTOI
values are unseeded single draws. The number of training epochs was selected on
held-out Corr2D, which can only favor the trained models over the controls.
Finally, our architecture is one point in a large design space. The controls
apply equally to a model that outperforms the template: they are what would make
that margin measurable.

\acknowledgments
This work is supported by NIH R01DC014717, R01DC018511 and R01CA133015.

\bibliography{main}
\bibliographystyle{spiebib}

\end{document}